\documentstyle[amsfonts,epsfig,12pt]{article}

\def\mypagenumber{1}

\def\myend{\end{document}}

\normalsize

\newcounter{sxn}

\newcounter{axn}

\date{}

\newdimen\mybaselineskip
\newcommand{\beeq}{\begin{equation}}
\newcommand{\eneq}{\end{equation}}
\newcommand{\be}{\begin{eqnarray}}
\newcommand{\ee}{\end{eqnarray}}
\newcommand{\bpic}{\begin{picture}}
\newcommand{\epic}{\end{picture}}

\def\la{\raise.16ex\hbox{$\langle$} \, }
\def\ra{\, \raise.16ex\hbox{$\rangle$} }

\def\psibar{ \psi \kern-.65em\raise.6em\hbox{$-$} }
\def\mbar{ m \kern-.78em\raise.4em\hbox{$-$}\lower.4em\hbox{} }

\def\L{ {\Lambda} }
\def\a{ {\alpha} }

\def\b{ {\beta} }
\def\L{ {\Lambda} }

\def\n@space{\nulldelimiterspace=0pt \mathsurround=0pt }
\def\huge#1{{\hbox{$\left#1\vbox to 20.5pt{}\right.\n@space$}}}

\def\myskip{\noalign{\kern 8pt}}
\def\myeqspace{\noalign{\kern 10pt}}

\def\boxit#1{$\vcenter{\hrule\hbox{\vrule\kern3pt
    \vbox{\kern3pt\hbox{#1}\kern3pt}\kern3pt\vrule}\hrule}$}
\def\bigbox#1{$\vcenter{\hrule\hbox{\vrule\kern5pt
     \vbox{\kern5pt\hbox{#1}\kern5pt}\kern5pt\vrule}\hrule}$}

\def\ignore#1{{}}

\begin{document}
\bibliographystyle{unsrt}
\footskip 1.0cm

\thispagestyle{empty}
\setcounter{page}{\mypagenumber}

             
\begin{flushright}{
BRX-TH-502\\}

\end{flushright}

\vspace{2.5cm}
\begin{center}
{\LARGE \bf { Gravitational Energy in Quadratic Curvature Gravities }}\\ 
\vskip 1 cm
{\large{S. Deser and Bayram Tekin  }}\footnote{e-mail:~
deser, tekin@brandeis.edu}\\
\vspace{.5cm}
{\it Department of  Physics, Brandeis University, Waltham, MA 02454,
USA}\\
\begin{abstract}
We define the notion of energy, and compute its values, for 
gravitational systems involving terms quadratic in curvature.  
While our construction parallels that of ordinary Einstein gravity, 
there are significant differences both conceptually and concretely.  
In particular, for $D=4$, all purely quadratic 
models admit vacua of arbitrary constant curvature. 
Their energies, including that of conformal (Weyl) gravity,
necessarily vanish in asymptotically flat spaces. Instead, they are 
proportional to that of the Abbott-Deser (AD) energy expression in 
constant curvature backgrounds and therefore also proportional 
to the mass parameter in the 
corresponding Schwarzschild-(Anti) de Sitter geometries. 
Combined Einstein-quadratic curvature systems reflect the above results: 
Absent a cosmological constant term,  the only vacuum is flat space, 
with the usual (ADM) energy and no explicit contributions from the
quadratic parts. If there is a $\Lambda$ term, then the vacuum is
 also unique with that $\Lambda$ value, and the energy 
is just the sum of the separate contributions from Einstein and quadratic 
parts to the AD expression . Finally, we discuss the effects on 
energy definition of both higher curvature terms and higher dimension.
\end{abstract}

\end{center}

\vspace*{1.5cm}

\newpage

\normalsize
\baselineskip=22pt plus 1pt minus 1pt
\parindent=25pt
\vskip 2 cm

General relativity is no different
from other effective low energy models, acquiring
higher momentum corrections, of quantum or string origin, 
to the Einstein action. They are represented 
locally by higher derivative additions. Coordinate invariance implies 
that these  local higher derivative additions consist 
of second and higher powers of curvatures (and their derivatives). 
The physics of
such ``improved'' models, as well as of those consisting of purely higher
derivative terms, is not immediately reducible to that of Einstein theory, 
just as addition of a cosmological (zero momentum) term profoundly changes 
the physics of the latter. In particular, the fundamental notion of 
gravitational energy is sufficiently different in presence of a $\Lambda$
term that more than two decades elapsed before its extension was obtained 
\cite{adm, abbott}. The next step in the momentum ladder likewise requires 
a clear understanding of {\it{its}} energy. Although quadratic curvature 
models, with or without Einstein or $\Lambda$ terms,
have long been studied, there has been a great deal of confusion about 
 their energy.  This has been due primarily to use of a flat, 
rather than the more relevant 
( even in the absence of a fundamental $\Lambda$ term, as we shall see ) 
constant curvature, vacuum. In this Letter, 
we intend to provide a universal definition of 
energy, and to evaluate it in appropriate asymptotic geometries, 
for theories quadratic (or higher) in curvatures, with or without 
Einstein and cosmological components. 

We will initially work in $D=4$, 
which, interestingly, is somewhat special. 
Since the Gauss-Bonnet invariant 
$\int d^4 x \sqrt {-g}( R_{\mu\nu\rho\sigma}^2 -4R^2_{\mu\nu}+R^2$), is
purely topological and does not contribute to field equations, the
generic quadratic curvature action is 
\be
  16\pi\kappa I = \int d^4\, x \sqrt{-g} (\a R^2 + \b R^2_{\mu\nu} ).
\label{action}
\ee
The famous conformal (Weyl) gravity corresponds to the choice $\b = -3\a$, 
but has no special energy features; instead, we will see that $\b = -4\a$ 
is unique in this respect. The ``post-Einstein'' constant $\kappa$ 
is of dimension $[GL^{-2}] = [M L]^{-1}$, leaving $(\a,\b)$ dimensionless.

Let us recall that there are two necessary facets of a proper 
energy definition: Firstly, identification of the ``Gauss law", whose
existence is guaranteed by gauge invariance; secondly choice of
the proper vacuum, possessing sufficient Killing symmetries with
respect to which global, background gauge invariant, generators
can be defined; these will always appear as surface integrals in the
asymptotic vacuum. Historically, the first application was of course to 
Einstein gravity without a cosmological term \cite{adm} whose natural vacuum
is flat space with its Poincar\'{e} symmetries.  The next
case, cosmological gravity, is a bit more involved \cite{abbott}, as
its vacua, de Sitter (dS) or anti-de Sitter (AdS) spaces 
have constant (rather than zero) curvature, necessarily
dictated by the cosmological constant $\L$ in the action.  The
relevant asymptotic symmetries are respectively $SO(4,1)$ and $SO(3,2)$,
which still support appropriate generators.  
Consider first the pure quadratic 
gravity of (\ref{action}).  Although the
equivalents of the Gauss law still exist and are still the ($0 \mu$)
components of the field equations, the choice of background is neither 
unique nor obvious. With conventions: 
signature $(-,+,+,+)$,
$[\nabla_\mu, \nabla_\nu]V_{\lambda} = 
R_{\mu \nu \lambda}\,^\sigma V_\sigma $,\,\, 
$ R_{\mu \nu} \equiv R_{\mu \lambda \nu}\,^\lambda$, 
the field equations are
\be
&&2\alpha R\,(R_{\mu\nu} - {1\over 4}g_{\mu\nu}\,R ) + 
(2\alpha +\beta) (g_{\mu\nu}\Box - \nabla_\mu \nabla_\nu )R \nonumber \\
&& +\beta \Box ( R_{\mu\nu} - {1\over 2}g_{\mu\nu} R)
+2\beta ( R_{\mu \sigma \nu \rho} - 
{1\over 4}g_{\mu \nu}R_{\sigma \rho})R^{\sigma \rho} = \kappa \tau_{\mu \nu},
\label{eom}
\ee
where we have introduced a (necessarily covariantly conserved) matter 
source $\tau_{\mu\nu}$. Now decompose the metric into the sum of a 
background $\bar{g}_{\mu \nu}$ 
(that solves the source-free version of (\ref{eom}) ) plus a deviation 
$h_{\mu \nu}$ of arbitrary strength,
\be
g_{\mu \nu} \equiv \bar{g}_{\mu \nu} + h_{\mu\nu}. 
\ee
As originally explained in \cite{abbott}, which we follow here, we 
separate the field equations into a part linear in $h_{\mu \nu}$ plus 
all the non-linear ones; the latter are moved to the right hand side, 
as part of the total source $T_{\mu\nu}$, that also includes $\tau_{\mu \nu}$, 
 thereby attaining the desired form
\be
{\cal{O}}(\bar{g})_{\mu \nu \alpha \beta}h^{\alpha \beta} = T_{\mu\nu}.
\label{ope}
\ee
The operator ${\cal{O}}(\bar{g})$ is hermitian and depends only on the
background metric (which also moves all indices and defines the covariant
derivatives $\bar{\nabla}_{\mu}$). It inherits both background Bianchi 
identity and (being hermitian) background gauge invariance from  
(the Bianchi identities of) the full theory, namely 
$\bar{\nabla}_{\mu}{\cal{O}}(\bar{g})^{\mu \nu \alpha \beta}=
{\cal{O}}(\bar{g})^{\mu \nu \alpha \beta}\bar{\nabla}_\a =0$.
As a consequence of these invariances, it is guaranteed 
that if the background $\bar{g}_{\mu\nu}$ is a vacuum 
that admits Killing vectors $\bar{\xi}_{\mu}$:
${\bar{\nabla}}_\mu \bar{\xi}_\nu +  {\bar{\nabla}}_\nu \bar{\xi}_\mu = 0$  , 
then there are associated conserved charges,
and they are expressible as surface integrals,
\be
Q^\mu(\bar{\xi}) = \int dS_i {\cal{F}}^{\mu i} 
\label{charge}
\ee
where ${\cal{F}}^{\mu \nu}$, an anti-symmetric tensor obtained from 
 ${\cal{O}}(\bar{g})$, depends on the specific model. This is easily 
verified by noting that $T_{\mu \nu}$ is both 
background-conserved and symmetric. 
Hence $\bar{\nabla}_{\mu} ( \sqrt{-\bar{g}} T^{\mu \nu}\bar{\xi}_\nu) \equiv  
\partial_\mu ( \sqrt{-\bar{g}} T^{\mu \nu}\bar{\xi}_\nu) = 0$, 
explicitly defining an 
ordinarily conserved vector current.  The energy is
simply that charge in (\ref{charge}) whose Killing vector is time-like.

It is here that the first departure from the Einstein framework
occurs: the theories of (\ref{action}) are scale-invariant 
and have no unique vacuum:  Any constant (or zero)
curvature space provides a candidate background.
For our systems (\ref{eom}), the detailed form of (\ref{ope}),  
about a constant curvature background reads
\be
T_{\mu\nu}= (2\alpha +\beta)({\bar{g}}_{\mu\nu}\bar{\Box} -
{\bar{\nabla}}_\mu {\bar{\nabla}}_\nu +\Lambda g_{\mu \nu})R_L
+4\Lambda (2\alpha + {\beta\over 3})\, {\cal{G}}^L_{\mu \nu} 
+\beta \bar{\Box} {\cal{G}}^L_{\mu\nu} - 
{2\beta\Lambda \over 3} \bar{g}_{\mu \nu}R_L,
\label{energymom}
\ee
where $\bar{\Box}= \bar{g}^{\mu \nu}\bar{\nabla}_\mu \bar{\nabla}_\nu $,  
${\cal{G}}^L_{\mu \nu} \equiv R^L_{\mu\nu} - 
{1\over 2}{\bar{g}}_{\mu\nu} R^L -\Lambda h_{\mu\nu}$ with 
$\bar{\nabla}^\mu {\cal{G}}^L_{\mu \nu}= 0$; we define 
$\bar{R}_{\mu\sigma\nu\rho} = {\Lambda \over 3}( \bar{g}_{\mu\nu} 
\bar{g}_{\sigma \rho} -  \bar{g}_{\mu\rho} 
\bar{g}_{\sigma \nu})$, so that 
$\bar{R}_{\mu\nu} = \Lambda \bar{g}_{\mu\nu}$. 
The degenerate, $\Lambda =0$, case $\bar{g}_{\mu \nu}= \eta_{\mu \nu}$ , 
just leads to $T_{\mu \nu} \rightarrow (\partial \partial R^L)_{\mu \nu}$, 
which necessarily implies that the energy of all asymptotically flat 
solutions of any purely quadratic model vanishes: 
this is an obvious aspect of the fact that equations of the form 
$\nabla^4 \phi = \rho$ 
are solved by $\phi \rightarrow r [ \int d^3x\, \rho] $: 
energy and source are not related by a Poisson operator. [This 
remark directly accounts for the well-known result \cite{boulware} 
that energy in Weyl gravity vanishes for asymptotically flat metrics,
but it is no different in this respect from any other $(\alpha, \beta)$ 
system.] 
We emphasize that while energy is too degenerate to be meaningful here, 
asymptotically flat solutions are not excluded thereby, 
nor do hamiltonian methods cease to be applicable, for example in 
the analysis of the excitation spectrum about flat space.  

We come next to the generic case of $\bar{g}_{\mu\nu}$ with $\Lambda \ne 0$. 
Here the linearization produces
a universal effect:  We find that   
\newpage
\be
8\pi\kappa Q^\mu(\bar{\xi}) &=& 
2\Lambda(4\alpha +\beta) \int d^3 x \sqrt{ -\bar{g}} 
\bar{\xi_\nu}{\cal{G}}_L^{\mu\nu}  \nonumber \\
&&+ (2\alpha +\beta) 
\int  dS_i \sqrt{-\bar{g}} \Big \{ \bar{\xi}^\mu 
\bar{\nabla}^i R_L + R_L \bar{\nabla}^\mu\, \bar{\xi}^i  
- \bar{\xi}^i \bar{\nabla}^\mu R_L \Big \} \nonumber \\
&& +\beta \int dS_i \sqrt{-\bar{g}}\Big \{ \bar{\xi}_\nu 
\bar{\nabla}^{i} {\cal{G}}_L^{\mu \nu}
- \bar{\xi}_\nu \bar{\nabla}^{\mu} {\cal{G}}_L^{i \nu} -
{\cal{G}}_L^{\mu\nu} 
\bar{\nabla}^{i} \bar{\xi}_\nu + {\cal{G}}_L^{i \nu} 
\bar{\nabla}^{\mu}\bar{\xi}_\nu \Big \}.
\label{fullcharge}
\ee
The integral in the first line  
is the standard charge of cosmological Einstein gravity \cite{abbott},
itself also a surface integral of course. [In obtaining the above,
gauge-invariant, surface form of the charge, it is helpful to organize 
the integrand to exhibit antisymmetry in $\mu$ and $i$.]  
Simple as this result is, it
becomes even nicer when we turn to the evaluation of 
the relevant, exterior, asymptotic solutions, namely the 
Schwarzschild-de Sitter and Schwarzschild-anti-de Sitter metrics;  
we label them collectively by SdS for brevity. [ Here, 
a major difference between dS and AdS,
that the former has an intrinsic horizon, enters. As
explained in \cite{abbott}, the dS energy definition is strictly
valid only inside the horizon, where the relevant Killing vector
stays time-like: This restriction also logically entails that the 
black hole horizon  be small compared to the cosmological one.  
We do not discuss the question of global definability or 
usefulness of dS energy
\cite{witten}, as it is really separate from the choice of dynamical
model.  No such problem affects the AdS case, where the surface
integrals may be taken at spatial infinity.] 
In cosmological Einstein gravity, 
$Q^0= \int \bar{\xi}_{\nu}{\cal{G}}_L^{0\nu}$ indeed gives the
desired value  $8\pi M G $, where $M$ is the
``Schwarzschild" mass in the SdS solutions.  Here, we find that the extra, 
second and 
third, lines of (\ref{fullcharge}) all vanish for  SdS 
spaces, so generically the energy is proportional to that of 
cosmological Einstein
gravity:
\be
E= 4\Lambda r_0  \kappa^{-1} ( 4\alpha +\beta), 
\ee
where $r_0$ is the coefficient of $1/r$ in the usual static form
of SdS, {\it{i.e}} the monopole moment of the total source, $\rho$,  
including (as 
always) gravitational contributions; it becomes proportional to the 
source mass, $m = \int d^3 x \tau_{00}$, for weak fields and sources, 
just as in Einstein theory. [More precisely, the Gauss 
equation is of the form $\Lambda \nabla^2 \phi = \kappa \rho$, so that $r_0
= ({\kappa/\Lambda})M$; the effective gravitational constant is
$({\kappa/ \Lambda})$ here.] Note that Weyl theory now has 
non-vanishing energy. Instead, it is the special  
$\beta = -4\alpha$ theory, whose action is the square of the traceless 
Ricci tensor $\tilde{R}_{\mu \nu} = R_{\mu\nu} -{1\over 4} g_{\mu \nu}R$ 
that has 
vanishing energy for all values of $\Lambda$.

We consider now the combined Einstein plus quadratic
curvature theories. If there is no explicit $\Lambda$ term, then
constant curvature spaces are no longer solutions of the combined
equations, and we are forced to flat background: consequently, the
Einstein term's energy expression is the whole story (which does
{\it not} mean that the quadratic terms do not contribute, as sources, 
to its value!). If instead, a $\Lambda$-term is also present, 
then constant curvature with precisely that $\Lambda$ value 
is not only an allowed, but {\it the} unique
vacuum; the scale is now fixed by the Einstein part, and the
energy is
\be
E = r_0 G^{-1} + 4\Lambda r_0 \kappa^{-1}  ( 4\alpha +\beta).
\ee
Here, $r_0$ is the ``Schwarzschild mass'' that solves the
Poisson equation, with contributions from both $R$ and $R^2$ parts.
For weak fields and sources, $E$ reassuringly becomes proportional to 
$m$, as before.
   
Thus far, we have worked in $D=4$ and considered only models 
with at most quadratic terms in curvatures. 
Depending on its physical origins, a given 
 higher power term may be viewed either as a part of the fundamental action 
( {\it{e.g.}}, if there is no Einstein term, as in Weyl gravity), 
or as a small correction that should not 
be in the ``kinetic'' term, nor a defining component of the energy 
expressions, though it still affects their values. However, even when viewed 
as sources, higher curvature powers pose a problem as they can 
prevent the existence of non-zero curvature background vacua. 
Put another way, their contributions to $T_{\mu \nu}$  
do not fall off at infinity. Take, for example, a 
generic higher curvature invariant $\int R^n$, $n> 2$, $R$
representing Riemann, Ricci or scalar curvatures, 
possibly involving also (an even number of) covariant
derivatives. Schematically, this gives rise to a field equation contribution
of the form $(R^n)_{\mu\nu} + (\nabla \nabla R^{n-1})_{\mu \nu}$. Just
as for $n=2$, its linearization about flat space does not affect the
energy. However generically these terms do not allow constant curvature 
solutions, since they are not homogeneous of order zero in the metric:
$\bar{R^n}_{\mu\nu}$ does not vanish even though 
$\bar{\nabla} \bar{\nabla} (\bar{R}^{n-1})_{\mu \nu}=0$ does. 
While the linearization, $\Lambda^{n-2}[ \Lambda +
\bar{\nabla} \bar{\nabla}] R_L$, does resemble that of
$n=2$, the background part $(\bar{R}^n)_{\mu\nu}$, acts (as mentioned above) 
as a constant source in the
Gauss equation. If we classify the nonlinear terms according to 
the powers of the three basis tensors: Weyl (traceless part of Riemann),
$\tilde{R}$ (traceless part of Ricci) and scalar curvature, then they will
allow cosmological vacua if they are not pure $R^n$. They also explicitly
contribute to the energy, as well as being sources of it, through the 
linear $h_{\mu \nu}$ terms in their field equations' expansions. 
                                        
The above complications arise immediately in $D > 4$: There, the quadratic
terms themselves are no longer  homogeneous of degree zero 
in the metric , seemingly forbidding (A)dS backgrounds.  There
are three independent invariants (unlike the two at $D=4$). We may take 
this basis to be (no longer conformally invariant!) Weyl gravity, the 
$\tilde{R}^2$ term and (scalar) $R^2$. The latter action forbids 
constant curvature, while the first two clearly permit it [see also
\cite{odintsov}] . Hence they also support our energy definition. 
The $\tilde{R}^2$ term's explicit contribution will vanish
just as it did in $D=4$. This leaves the Weyl gravity as both
allowing nonflat vacuum and contributing explicitly to energy. The 
relevant asymptotic metric here is again SdS (or slightly generalized 
form allowed for pure Weyl theory). If Einstein and cosmological terms 
are also present, our earlier $D=4$ analysis applies straightforwardly. 

In summary, we have defined energy for arbitrary general covariant
gravitational models, particularly the simplest, quadratic
curvature, systems. In $D=4$, pure quadratic actions have useful, 
non-vanishing energy (only) with respect to cosmological backgrounds.  
While these vacua are infinitely degenerate, 
their value for any chosen $\Lambda$ is quite
physical, being proportional to that of the Schwarzschild mass in
the relevant SdS metric. Models with both Einstein and quadratic 
actions differ in imposing a unique background; here
the total energy is the sum of contributions proportional to the
cosmological (Abbott-Deser) mass, if there is also an explicit 
$\Lambda$-term, and just equal to the ADM mass if $\Lambda=0$ .
We also studied the effect on energy definition of higher curvature 
invariants and its extension to $D > 4$. The general framework could be
maintained if the higher powers contained curvature
combinations such as Weyl tensors that vanish in (A)dS, 
and not just the scalar curvature. For $D >4$, there are three 
independent quadratic terms: If $R^2$ is present, flat vacuum 
is forced. The other two combinations, in particular Weyl 
gravity, allow (A)dS backgrounds and hence permit $D=4$ energy construction. 
Details will be presented elsewhere.

This work was supported by National Science Foundation Grant PHY99-73935.

\myend
\begin{thebibliography}{99}

\bibitem{adm}
R.~Arnowitt, S.~Deser and C.~Misner, Phys. \ Rev.\ {\bf 116}, 1322 (1959);
{\bf 117}, 1595 (1960); in {\it{Gravitation: an introduction to current
research}}, ed L. Witten (Wiley, New York, 1962)


\bibitem{abbott}
L.~F.~Abbott and S.~Deser,
Nucl.\ Phys.\ B {\bf 195}, 76 (1982).


\bibitem{boulware}
D.~G.~Boulware, G.~T.~Horowitz and A.~Strominger,
Phys.\ Rev.\ Lett.\  {\bf 50}, 1726 (1983).

\bibitem{witten}
E.~Witten,
``Quantum gravity in de Sitter space,''
arXiv:hep-th/0106109.
A.~Strominger,
JHEP {\bf 0110}, 034 (2001);
 

\bibitem{odintsov}
M.~Cvetic, S.~Nojiri and S.~D.~Odintsov,
Nucl.\ Phys.\ B {\bf 628}, 295 (2002)
[arXiv:hep-th/0112045].

\end{thebibliography}
